\documentclass[sigconf]{acmart}

\copyrightyear{2026}
\acmYear{2026}
\setcopyright{cc}
\setcctype{by}
\acmConference[ICSE-Companion '26]{2026 IEEE/ACM 48th International Conference on Software Engineering}{April 12--18, 2026}{Rio de Janeiro, Brazil}
\acmBooktitle{2026 IEEE/ACM 48th International Conference on Software Engineering (ICSE-Companion '26), April 12--18, 2026, Rio de Janeiro, Brazil}
\acmPrice{}
\acmDOI{10.1145/3774748.3787746}
\acmISBN{979-8-4007-2296-7/2026/04}

\usepackage{acronym}
\acrodef{LLM}{Large Language Model}

\usepackage{makecell}

\begin{document}

\title{\textsc{AgenticTyper}: Automated Typing of Legacy Software Projects Using Agentic AI}

\author{Clemens Pohle}
\orcid{0009-0007-5507-9827}
\affiliation{%
  \institution{Darmstadt University of Applied Sciences}
  \city{Darmstadt}
  \country{Germany}
}
\email{clemens.pohle@stud.h-da.de}
\affiliation{%
  \institution{MaibornWolff GmbH}
  \city{Darmstadt}
  \country{Germany}
}
\email{clemens.pohle@maibornwolff.de}

\renewcommand{\shortauthors}{Clemens Pohle}

\begin{abstract}
  Legacy JavaScript systems lack type safety, making maintenance risky.
  While TypeScript can help, manually adding types is expensive.
  Previous automated typing research focuses on type inference but rarely addresses type checking setup, definition generation, bug identification, or behavioral correctness at repository scale.
  We present \textsc{AgenticTyper}, a \ac{LLM}-based agentic system that addresses these gaps through iterative error correction and behavior preservation via transpilation comparison.
  Evaluation on two proprietary repositories (81K LOC) shows that \textsc{AgenticTyper} resolves all 633 initial type errors in 20 minutes, reducing manual effort from one working day.
\end{abstract}



\maketitle

\section{Problem and Motivation}

Business-critical legacy systems, often written in dynamically typed languages like JavaScript~\cite{jetbrains2025,stackexchange2025}, accumulate significant business value but become increasingly difficult to modify and understand~\cite{assuncao2025,khadka2014}.
Lack of type safety makes changes risky, potentially breaking existing functionality~\cite{feathers2004}.
Optional typing systems~\cite{bracha2004} like TypeScript provide documentation~\cite{hanenberg2014}, make dependencies explicit~\cite{jin2023}, and prevent bugs~\cite{gao2017}.
However, manually adding type safety is costly~\cite{gao2017}, often requiring years~\cite{yee2023}.
Developers must trace dependencies across files, design reusable type definitions, and address cascading issues as each type annotation may reveal new problems in dependent code.
Therefore, we present ongoing work on \textsc{AgenticTyper}\footnote{Source code: \url{https://github.com/clemens-mw/agentic-typer}}~\cite{pohle2025}, leveraging \acf{LLM}-based coding agents to automate typing legacy JavaScript repositories without regressions.

\section{Background and Related Work}

Automated typing research focuses on type inference using static~\cite{aiken1991,jensen2009}, dynamic~\cite{an2011,pizzorno2025}, probabilistic~\cite{raychev2015,wei2019}, and hybrid methods~\cite{pradel2020,pandi2021}.
However, these approaches face key limitations~\cite{guo2024}: static methods cannot handle dynamic features~\cite{sun2023}, dynamic methods require test suites~\cite{hoeflich2022}, and probabilistic methods produce incorrect predictions that must be discarded~\cite{pradel2020}.
Generating type definitions~\cite{yee2024} and identifying~\cite{pradel2015} and fixing~\cite{oh2022} existing type issues remain underexplored.

\acp{LLM} have been applied to numerous software engineering tasks~\cite{hou2024}, including type inference~\cite{jesse2021,jesse2023,cassano2023}, fixing type errors~\cite{chow2024,peng2024,hao2024}, and combined with static analysis~\cite{seidel2023,bharti2025}.
However, simply prompting an \ac{LLM} to add types fails at repository scale: models cannot navigate and edit multiple files, handle cascading errors, or verify behavioral preservation.
\ac{LLM}-based agents~\cite{xi2025} (\textit{agentic AI}~\cite{kapoor2025}) address this through tool use and multi-step reasoning and have recently been applied to program repair~\cite{bouzenia2025,joos2025}, but not yet to typing tasks.

\section{Approach and Uniqueness}

Unlike prior work, \textsc{AgenticTyper} combines agentic AI with transpilation comparison for change detection, enabling repository-level typing without test suites.
Additionally, it distinguishes between actual bugs requiring human attention and valid patterns that cannot be typed without behavioral changes (e.g., type coercion).

\begin{figure}
\includegraphics[width=\columnwidth]{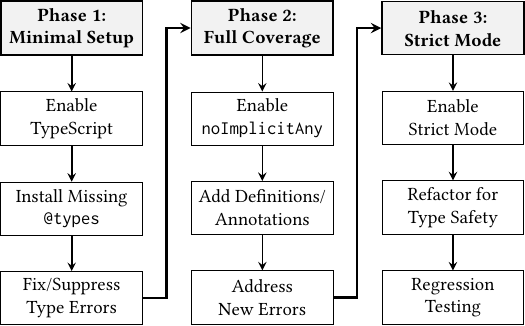}
\caption{Three-phase typing approach.}
\label{fig:phases}
\Description{A flowchart diagram showing three sequential phases for automated typing. Phase 1 (Minimal Setup) includes enabling TypeScript, installing missing types packages, and fixing or suppressing type errors. Phase 2 (Full Coverage) enables noImplicitAny, adds definitions and annotations, and addresses new errors. Phase 3 (Strict Mode) enables strict mode, refactors for type safety, and performs regression testing. Thick arrows connect the phases sequentially.}
\end{figure}

We propose three incremental phases (Figure~\ref{fig:phases}).
\textbf{Phase one} establishes minimal TypeScript checking.
Coding agents analyze and address type errors (TypeScript compiler diagnostics) in each file by adding or fixing JSDoc type annotations (e.g., \texttt{@type}, \texttt{@param}).
When errors cannot be resolved without behavioral changes, agents insert \textit{suppression comments} (\texttt{@ts-expect-error}): bugs are marked for human review, while valid patterns receive neutral explanations, creating a clean baseline.
\textbf{Phase two} enables \texttt{noImplicitAny}, requiring explicit annotations throughout.
Agents add type definitions and annotations, addressing newly surfaced errors as in phase one.
\textbf{Phase three} enables strict checking, revealing issues that often require refactoring (e.g., null checks).
Since refactoring carries risk, this phase requires regression testing via generated tests or manual verification.

\begin{figure}
\includegraphics{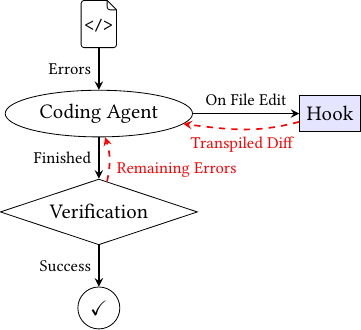}
\caption{Single-agent architecture for phase one and two. The hook preventing behavior changes is highlighted.}
\label{fig:agent}
\Description{A flowchart showing the workflow of a single agent. Starting from a code file at the top, errors flow to a coding agent. A hook triggers on file edits and sends back transpiled diff via a red dashed feedback arrow if there are any changes. When the agent finishes, a verification step is executed. If remaining errors exist, a red dashed feedback arrow returns to the agent. On success, the flow completes with a checkmark.}
\end{figure}

In phases one and two, the behavior preservation mechanism (Figure~\ref{fig:agent}) prevents regressions without tests: after each edit, a hook transpiles the modified code to JavaScript and compares it against the original transpiled output, alerting agents of any runtime changes.
After an agent finishes, static verification checks for remaining errors before proceeding.
Multiple agents run in parallel, and a final verification agent addresses cross-file errors introduced by earlier fixes.

\section{Results and Contributions}

So far, we implemented phase one using the Claude Agent SDK~\cite{claude2025} with Claude Sonnet 4.5~\cite{sonnet2025}, chosen for its leading coding capabilities as of September 2025.
We evaluated \textsc{AgenticTyper} on two proprietary legacy Node.js backend repositories from an industrial facility management platform (Table~\ref{tab:results}).
The system is configured to run ten agents in parallel.

\begin{table}
\centering
\caption{Phase one results on two proprietary JavaScript repositories. Average of three runs each.}
\label{tab:results}
\resizebox{\columnwidth}{!}{%
\begin{tabular}{lrrrrrrr}
\toprule
\textbf{Repo} & \textbf{LOC} & \makecell{\textbf{Type}\\\textbf{Errors}} & \makecell{\textbf{Necessary}\\\textbf{Suppressions}} & \makecell{\textbf{Additional}\\\textbf{Suppressions}} & \textbf{Time} & \textbf{Cost} \\
\midrule
A & 75K & 570 & 327 & +26 (+8.0\%) & 16:51 min & \$22.85 \\
B & 6K  & 63  & 56  & +0  (+0.0\%) & 3:06 min  & \$2.08 \\
\midrule
\textbf{Total} & \textbf{81K} & \textbf{633} & \textbf{383} & \textbf{+26 (+6.8\%)} & \textbf{19:57 min} & \textbf{\$24.93} \\
\bottomrule
\end{tabular}%
}
\end{table}

\textsc{AgenticTyper} successfully establishes type checking and addresses all 633 initial type errors without modifying runtime behavior, verified through transpilation comparison.
This task previously required one full working day by an experienced TypeScript developer and is now completed in 20 minutes for \$25.
While \textsc{AgenticTyper} handles most cross-file dependencies properly, 7\% of suppressions could have been avoided with more precise root cause tracing.
Examples include parameter shadowing and bugs where the agent suppressed at multiple usage sites rather than the single source.

In summary, our contributions are: (1) a novel agentic approach with behavior preservation for automated typing, (2) an open-source implementation with parallel agent execution, and (3) an evaluation demonstrating practical feasibility and cost-effectiveness.

\section{Future Work}

We will implement and evaluate phases two and three on the same repositories, plus a 300,000 LOC TypeScript repository where phase one is not necessary.
Evaluation will measure type coverage (percentage of typed elements), annotation correctness through type checking and manual review, and cleanup effort in hours and changed lines.
We will also investigate different \acp{LLM} and extend to other optionally typed languages like Python.

\balance
\bibliographystyle{ACM-Reference-Format}
\bibliography{bibliography}

\end{document}